\documentclass[twocolumn]{aastex61}

\received{July 1, 2016}
\revised{September 27, 2016}
\accepted{\today}
\submitjournal{ApJ}

\shorttitle{Abundances in NGC 6791}
\shortauthors{Villanova et al.}

\begin{document}

\title{NGC 6791: a probable bulge cluster without multiple populations.}
       \thanks{Based on observations carried out at La Silla Paranal Observatory
under program 095.D-0294.}

\correspondingauthor{Sandro Villanova}
\email{svillanova@astro-udec.cl}

\author{Sandro Villanova}
\affiliation{Departamento de Astronomia, Casilla 160-C, Universidad de Concepcion, Chile}

\author{Giovanni Carraro}
\affiliation{Dipartimento di Fisica e Astronomia {\it Galileo Galilei}, Universit\'a di Padova,
Vicolo Osservatorio 3, I-35122, Padova, Italy}

\author{Doug Geisler}
\affiliation{Departamento de Astronomia, Casilla 160-C, Universidad de
  Concepcion, Chile}
\affiliation{Instituto de Investigación Multidisciplinario en Ciencia y
  Tecnología, Universidad de La Serena. Avenida Raúl Bitrán S/N, La Serena,
  Chile}
\affiliation{Departamento de Física y Astronomía, Facultad de Ciencias,
  Universidad de La Serena. Av. Juan Cisternas 1200, La Serena, Chile}
\nocollaboration

\author{Lorenzo Monaco}
\affiliation{Departamento de Ciencias Fisicas, Universidad Andres Bello,
Fernandez Concha 700, Las Condes, Santiago, Chile}
\nocollaboration

\author{Paulina Assmann}
\affiliation{Departamento de Astronomia, Casilla 160-C, Universidad de Concepcion, Chile}
\nocollaboration



\begin{abstract}
NGC 6791 is a unique stellar cluster, key to our understanding of both the multiple stellar population phenomenon
and the evolution and assembly of the Galaxy. However, despite many investigations, its nature is still very controversial. \citet{Gei12} found evidence suggesting it was the first open cluster to possess multiple populations but several subsequent studies did not corroborate this.
It has also been considered a member of the thin or thick disk or even the bulge,
and both as an open or globular cluster or even the remnant of a dwarf galaxy.

Here, we present and discuss detailed abundances derived from high resolution spectra obtained with
UVES at VLT and HIRES at Keck of 17 evolved stars of this cluster. We obtained a mean [Fe/H]=$+0.313\pm$0.005, 
in good agreement with recent estimates, and with no indication of star-to-star metallicity variation, as expected. 
We also did not find any variation in Na, in spite of having selected the very same stars as in \citet{Gei12}, where a Na variation was claimed.
This points to the presence of probable systematics in the lower resolution spectra of this very high metallicity cluster analysed in that work. In fact, we find no
evidence for an intrinsic spread in any element, corroborating recent independent APOGEE data. 
The derived abundances indicate  that NGC 6791 very likely formed in the  Galactic Bulge and that the  proposed association with the Thick Disk is unlikely, despite its present Galactic location. We confirm the most recent hypothesis suggesting that the cluster could have formed in the Bulge and radially migrated to its current location, which appears the best explanation for this intriguing object.
\end{abstract}




\section{Introduction} \label{sec:intro}

NGC 6791 is a remarkable, fascinating  Milky Way star cluster. From the first detailed studies \citep{Kin65,Spi71}, its properties were recognized as being extreme. 
Its combination of very old age and very high metallicity ($\sim$8 Gyr and  [Fe/H] = 0.3$-$0.4; \citet{Cu15}\ and references therein) is in fact unique among open clusters in our Galaxy. 
In addition, in spite of its old age and location at the solar Galactocentric distance,
this cluster is still one of the most massive ($M\sim 10^4 M_{\odot}$) old open
clusters  known in the Galaxy.  Indeed, there is evidence that it has undergone substantial mass loss via tidal interactions \citep{Da15} and was therefore much more massive in the past. Another mystery surrounding its nature is how did such a high metallicity object reach its current location 8kpc from the Galactic center and a kpc from the plane? It is an infamous outlier in all age-metallicity relations of Galactic disk objects (e.g. \citet{Net16}). 

The nature and origin of this cluster is very
controversial. An impressive variety of scenarios have been suggested, including 
the possibility that it is a Thick Disk cluster \citep{Lin17}, an 
extragalactic,  strongly mass-depleted dwarf elliptical \citep{Car06}, or a
Bulge/Inner Disk star cluster \citep{Jil12, Mar17}.

Several recent spectroscopic studies
have revealed stronger and stronger hints that NGC 6791 was chemically anomalous as well. \citet{Huf95} found evidence for CN (but not CH) variations in a number of red clump (RC) stars measured using low resolution spectra, reminiscent of the first signs of multiple populations seen in globular clusters long ago \citep{Hes77}. Similar behavior was found for stars covering a range of evolutionary status including the MS, RGB and RC, again from low resolution spectra from the SEGUE survey \citep{Car12}. The  detailed high resolution study by \citep{Gei12} suggested that NGC 6791 actually harbors
multiple stellar populations, based on the detection of two groups of evolved stars with significantly different Na abundance. This would make NGC 6791 the first open cluster to possess multiple populations, which until now have been limited to more massive globular clusters \citep{Car09,Muc16}. 
Together with the suggestion that star formation could have lasted as long as 1 Gyr in the cluster \citep{Tw11}, this would  make NGC 6791 more similar to Galactic globular or Magellanic Cloud massive clusters, where Na variations or extended star formation histories
are routinely found \citep{Car10, Bau07}.

However, more recent spectroscopic studies, both high and low resolution 
\citep{Bra14,Boe15,Cu15,Bob16,Lin17} have not detected any indication of multiple stellar populations. This clearly casts doubt on the reliability of the previous results.

Obviously, our knowledge of the nature, origin and detailed characteristics of this 
unique object is sorely lacking. One would like to pin down the origin of such an exotic object, definitively determine to which Galactic component it belongs, whether or not it exhibits multiple populations and its relationship to the Galactic Globular and Open cluster population. This is the main motivation for the present study.
A proper assessment of the cluster chemical characteristics, nature and origin is still
missing, and here we revisit this topic again to try and shed new light on its many mysteries.

To this aim, we collected high resolution (R=47,000) VLT UVES spectra of 14 giants. In order to control uncertainties as much as possible,
we selected the very same stars for which previous WIYN Hydra spectra indicated Na
abundance variations \citep{Gei12}. In addition, we reanalyzed the high-resolution
Keck HIRES spectra reported in that paper. Our main goal is to investigate whether our previous results indicating the presence of multiple populations are confirmed or denied.
The new abundances also allow us to place the cluster into the larger Galactic
context, and provide additional information concerning its origin. 
The paper layout is as follows. In section 2 we introduce our new
observational material and the data reduction, in section 3 we describe the
abundance analysis, in section 4 we present our results, and in section 5 we
discuss the implications of our results and give our conclusions.

\begin{figure}
\plotone{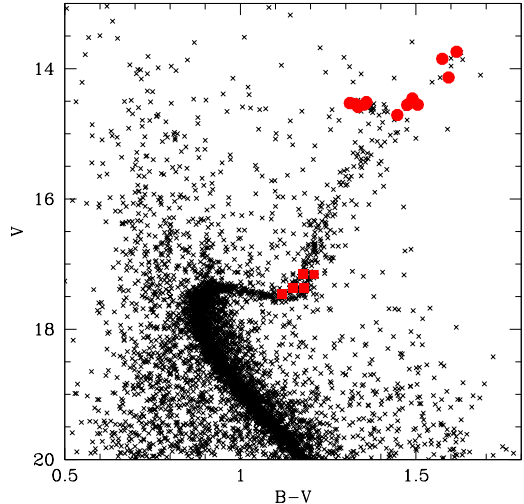}
\caption{CMD of NGC 6791. HIRES  targets are indicated with red filled
  squares, UVES targets with red filled circles.}
\label{fig1}
\end{figure}

\section{Observations and data reduction}

Our data consists of two sets of high resolution spectra. The first are
the same HIRES spectra collected at the Keck telescope and first analyzed in
\citep{Gei12}. The 5 stars lie on the reddest extension of the SGB/faintest extension of the RGB (see Fig.\ref{fig1}).
The second set are data collected at the FLAMES@UVES
spectrograph mounted at the VLT-UT2 telescope under the program
095.D-0294(A). In this case, targets are  14 of the upper RGB and RC stars observed previously in \citet{Gei12} with the HYDRA instrument on the WIYN telescope. These stars were observed with two fiber configurations.
We used the 580nm set-up, that gives a spectral coverage between
4800 and 6800~\AA\ with a resolution of R=47000. This resolution is more than three times higher than the resolution of our previous HYDRA spectra and the wavelength coverage is many times greater, providing many more lines. The signal-to-noise (S/N) ratio was
between 40 and 50 at 6000~\AA. 
Data were reduced using the dedicated pipeline version 5.9.1\footnote{see http://www.eso.org/sci/software/pipelines/}.
Data reduction includes bias subtraction, flat-field correction, wavelength calibration,
sky subtraction, and spectral rectification.
Radial velocities were measured using the {\it fxcor} package in IRAF \footnote{IRAF is distributed by the National
Optical Astronomy Observatory, which is operated by the Association of
Universities for Research in Astronomy, Inc., under cooperative agreement
with the National Science Foundation, see \citet{Tod93}}, using a synthetic spectrum as a template. 
The cluster turns out to have a radial velocity of -47.0$\pm$0.4 km/s and a velocity dispersion of 1.5$\pm$0.3 km/s.
All stars were confirmed to be cluster members with the exception of T04, for
which we measured a radial velocity of -60.8 km/s. This could indicate a
binary nature for this star since in \citet{Gei12} it has a radial
velocity compatible with the mean value of the cluster.
However we prefer to leave it out of the current analysis.
We left out of the analysis also the star T19. Its radial velocity of -47.41 km/s is
compatible with the cluster mean value but, because of the low temperature ($\sim$3800 K), its spectrum shows strong blending between atomic and  molecular lines and a continuum level that is very difficult to be determined. 
Table \ref{t1} lists the basic parameters of the 12 remaining stars:
ID from Stetson's photometry \citep{Ste03}, ID from \citet{Gei12}, J2000.0 coordinates (RA \& DEC in degrees), heliocentric radial velocity RV$_{\rm H}$ (km/s), and B,I,V, J, H, K$_s$ magnitudes.

\begin{table*}
\tabcolsep 0.05truecm
\caption{ID, coordinates, radial velocities and magnitudes of the observed stars. The typical error in radial velocity is 0.5 km/sec. G12 indicate Geisler et al. (2012) identification. See text for details.}
\begin{tabular}{lcccccccccc}
\hline
ID &ID(G12) & RA(2000.0) &DEC(2000.0) & $V_R$ &B &V &I &J &H &K\\
\hline
 & & [deg] & [deg] & [km/s] & mag & mag & mag & mag & mag & mag \\
\hline\hline
12630 &T06 &290.15787500 &37.74702778 &-47.89 &15.841 &14.529 &13.168 &12.222 &11.606 &11.435\\
13579 &T14 &290.17020833 &37.77275000 &-48.60 &15.904 &14.551 &13.217 &12.108 &11.503 &11.344\\
14235 &T18 &290.17804167 &37.85213889 &-47.40 &15.874 &14.515 &13.176 &12.248 &11.648 &11.488\\ 
14537 &T09 &290.18154167 &37.78391667 &-49.57 &16.160 &14.713 &13.230 &12.174 &11.503 &11.325\\
16927 &T03 &290.20695833 &37.73555556 &-48.11 &15.923 &14.588 &13.264 &12.273 &11.685 &11.500\\
18113 &T15 &290.21916667 &37.74125000 &-44.72 &15.729 &14.136 &12.373 &11.135 &10.417 &10.185\\
18243 &T05 &290.22037500 &37.75927778 &-45.21 &15.874 &14.546 &13.235 &12.280 &11.669 &11.513\\
18444 &T07 &290.22245833 &37.80788889 &-47.34 &15.357 &13.741 &11.962 &10.732 &9.962  &9.769\\
18772 &T17 &290.22579167 &37.77466667 &-45.18 &16.059 &14.554 &12.988 &11.857 &11.170 &10.978\\
19234 &T12 &290.23045833 &37.72100000 &-47.16 &16.032 &14.557 &13.027 &11.945 &11.269 &11.088\\
21447 &T11 &290.25470833 &37.70383333 &-46.70 &15.949 &14.459 &12.890 &11.821 &11.130 &10.938\\
22559 &T10 &290.26779167 &37.78858333 &-46.43 &15.424 &13.849 &12.191 &11.014 &10.310 &10.102\\
08506 &T31 &290.22495833 &37.77830556 &-46.30 &18.329 &17.150 &15.954 &14.770 &14.332 &14.335\\
09609 &T32 &290.23833333 &37.79583333 &-38.80 &18.368 &17.158 &15.923 &15.128 &14.512 &14.357\\
11014 &T33 &290.25620833 &37.74683333 &-44.80 &18.575 &17.457 &16.330 &15.577 &15.210 &14.904\\
11092 &T34 &290.25733333 &37.77563889 &-46.30 &18.553 &17.372 &16.164 &15.394 &14.727 &14.623\\
12383 &T35 &290.27687500 &37.76058333 &-48.00 &18.520 &17.370 &16.210 &  -    &  -    &  -   \\
\hline\hline
\end{tabular}
\label{t1}
\end{table*}

\begin{table*}
\tabcolsep 0.05truecm
\caption{Parameters and abundances of the observed stars. The last row gives the
  mean abundances of the cluster and the relative error of the mean.}
\begin{tabular}{lccccccccccccccc}
\hline
ID & $T_{eff}$ &log(g) & $v_t$ &[Fe/H]&[Na/Fe]&[Mg/Fe]&[Al/Fe]&[SiFe]&[Ca/Fe]&[Ti/Fe]&[Cr/Fe]&[Ni/Fe]&[Y/Fe]&[Ba/Fe]&[Eu/Fe]\\
\hline
& $^oK$ & &[km/s] & dex & dex & dex & dex & dex & dex & dex & dex & dex & dex & dex & dex \\ 
\hline\hline
12630 & 4376 & 1.99 & 1.31 & 0.28 & -0.05 & 0.24 & 0.20 &   -   & -0.06 & 0.09 & 0.07 & 0.16 &  0.22 &  0.06 & 0.10\\    
13579 & 4402 & 2.02 & 1.31 & 0.31 & -0.02 & 0.12 & 0.17 & -0.01 & -0.07 & 0.06 & 0.11 & 0.12 &  0.06 & -0.06 & 0.12\\  
14235 & 4444 & 2.03 & 1.35 & 0.32 & -0.15 & 0.18 & 0.21 & -0.08 & -0.15 & 0.25 & 0.12 & 0.20 &  0.07 &  0.08 & 0.22\\  
14537 & 4201 & 1.96 & 1.13 & 0.30 & -0.08 & 0.12 & 0.32 &  0.00 &  0.00 & 0.20 & 0.15 & - -- &  0.16 &  0.06 & 0.17\\    
16927 & 4405 & 2.03 & 1.30 & 0.34 & -0.06 & 0.12 & 0.26 & -0.03 & -0.04 & 0.06 & 0.06 &  -   &  0.09 &  0.02 & 0.20\\   
18113 & 3918 & 1.44 & 1.25 & 0.31 & -0.07 & 0.14 & 0.32 &   -   &  0.04 & 0.18 &  -   & 0.19 &  0.27 &  0.00 & 0.27\\    
18243 & 4441 & 2.04 & 1.34 & 0.31 & -0.06 & 0.13 & 0.25 &  0.03 & -0.06 & 0.09 & 0.07 & 0.09 &  0.12 & -0.07 & 0.12\\   
18444 & 3926 & 1.27 & 1.42 & 0.34 & -0.07 & 0.17 & 0.15 &  0.00 &   -   & 0.16 & 0.14 &  -   &  0.03 &  0.07 & 0.10\\     
18772 & 4095 & 1.80 & 1.14 & 0.34 & -0.09 & 0.16 & 0.22 &  0.11 & -0.07 & 0.23 & 0.08 & 0.13 &  0.18 &  0.11 & 0.15\\    
19234 & 4160 & 1.86 & 1.17 & 0.34 & -0.06 & 0.17 & 0.29 & -0.08 & -0.06 & 0.09 & 0.23 & 0.08 &  0.07 &  0.08 & 0.24\\   
21447 & 4139 & 1.80 & 1.19 & 0.32 & -0.04 & 0.17 & 0.27 &  0.01 & -0.01 &  -   &  -   & 0.20 &  0.10 &  0.00 & 0.15\\  
22559 & 4009 & 1.40 & 1.40 & 0.29 & -0.15 & 0.14 & 0.22 &  0.04 & -0.07 & 0.12 & 0.12 & 0.09 & -0.05 &  0.04 & 0.14\\  
08506 & 4674 & 3.26 & 0.47 & 0.31 & -0.11 & 0.13 &  -   &  0.03 &   -   & 0.15 & 0.09 & 0.14 &  0.16 &  0.01 &  -  \\     
09609 & 4647 & 3.25 & 0.45 & 0.31 & -0.06 & 0.22 &  -   &  0.02 & -0.13 & 0.11 & 0.09 & 0.24 &  0.12 & -0.07 &  -  \\   
11014 & 4869 & 3.49 & 0.50 & 0.29 & -0.05 & 0.12 &  -   &  0.01 &   -   & 0.06 & 0.09 & 0.16 &  0.11 &  0.07 &  -  \\     
11092 & 4702 & 3.37 & 0.41 & 0.29 & -0.07 & 0.10 &  -   &  0.03 & -0.05 & 0.10 & 0.09 &  -   &  0.11 &  0.05 &  -  \\    
12383 & 4775 & 3.40 & 0.46 & 0.33 & -0.08 & 0.16 &  -   & -0.03 &   -   &  -   & 0.10 &  -   &  0.09 &  0.06 &  -  \\   
\hline
Mean  &    &       &      & 0.314& -0.07 & 0.15 & 0.24 &  0.00 & -0.06 & 0.13 & 0.11 & 0.15 &  0.11 &  0.03 & 0.17\\
Error &    &       &      & 0.005&  0.01 & 0.01 & 0.02 &  0.01 &  0.01 & 0.02 & 0.01 & 0.01 &  0.02 &  0.01 & 0.02\\
\end{tabular}
\label{t2}
\end{table*}

\section{Abundance analysis}

\begin{table*}
\tabcolsep 0.05truecm
\caption{Estimated errors on abundances due to errors on atmospheric
parameters and to spectral noise for star $\#18243$ (column 2 to 6). Column 7
gives the total error calculated as the square root  of the sum of the squares
of columns 2 to 6. This total error must be compared with the total error as
obtained from the observed dispersion (RMS) of the data with its error (column
8). The last column gives the significance of the difference between
the total error for star $\#18243$ and the observed dispersion, in units of $\sigma$. Values within brackets are those calculated using the more conservative errors on the parameters (see text).}                         
\begin{tabular}{lccccccccc}        
\hline\hline  
El. & \footnotesize{$\Delta$T$_{\rm eff}$=10(50) K}  & \footnotesize{$\Delta$log(g)=0.05(0.20)} & \footnotesize{$\Delta$v$_{\rm t}$=0.04(0.10) km/s} & \footnotesize{$\Delta$[Fe/H]=0.01(0.05)} & S/N & $\Delta_{\rm tot}$ & $RMS_{\rm obs}$ & Sgn. ($\sigma$)\\
\hline
$\Delta$([Na/Fe]) & 0.01(0.05) & 0.01(0.03) & 0.01(0.03) & 0.00(0.01) & 0.05 & 0.05(0.08) & 0.04$\pm$0.01 & 1(4)\\
$\Delta$([Mg/Fe]) & 0.00(0.03) & 0.02(0.07) & 0.00(0.00) & 0.02(0.06) & 0.04 & 0.05(0.10) & 0.04$\pm$0.01 & 1(6)\\
$\Delta$([Al/Fe]) & 0.01(0.04) & 0.01(0.03) & 0.00(0.01) & 0.00(0.01) & 0.07 & 0.07(0.09) & 0.05$\pm$0.01 & 2(4)\\
$\Delta$([Si/Fe]) & 0.01(0.05) & 0.01(0.04) & 0.01(0.02) & 0.00(0.01) & 0.05 & 0.05(0.08) & 0.05$\pm$0.01 & 0(3)\\
$\Delta$([Ca/Fe]) & 0.01(0.04) & 0.01(0.03) & 0.00(0.01) & 0.00(0.01) & 0.04 & 0.04(0.07) & 0.05$\pm$0.01 & 1(2)\\
$\Delta$([Ti/Fe]) & 0.01(0.05) & 0.02(0.05) & 0.01(0.03) & 0.00(0.02) & 0.05 & 0.06(0.09) & 0.06$\pm$0.01 & 1(3)\\
$\Delta$([Cr/Fe]) & 0.01(0.05) & 0.01(0.03) & 0.00(0.02) & 0.00(0.01) & 0.05 & 0.05(0.08) & 0.04$\pm$0.01 & 0(4)\\
$\Delta$([Fe/H])  & 0.00(0.03) & 0.01(0.03) & 0.02(0.03) & 0.00(0.01) & 0.02 & 0.03(0.06) & 0.020$\pm$0.003 & 3(13)\\
$\Delta$([Ni/Fe]) & 0.00(0.02) & 0.01(0.03) & 0.00(0.00) & 0.00(0.01) & 0.06 & 0.06(0.07) & 0.05$\pm$0.01 & 1(2)\\
$\Delta$([Y/Fe])  & 0.00(0.04) & 0.02(0.06) & 0.03(0.07) & 0.00(0.02) & 0.06 & 0.07(0.12) & 0.07$\pm$0.01 & 1(5)\\
$\Delta$([Ba/Fe]) & 0.00(0.03) & 0.02(0.08) & 0.04(0.09) & 0.01(0.04) & 0.05 & 0.07(0.14) & 0.05$\pm$0.01 & 1(9)\\
$\Delta$([Eu/Fe]) & 0.00(0.03) & 0.02(0.07) & 0.01(0.02) & 0.00(0.02) & 0.08 & 0.08(0.11) & 0.06$\pm$0.01 & 2(5)\\
\hline                                   
\end{tabular}
\label{t3}
\end{table*}

\begin{figure}
\plotone{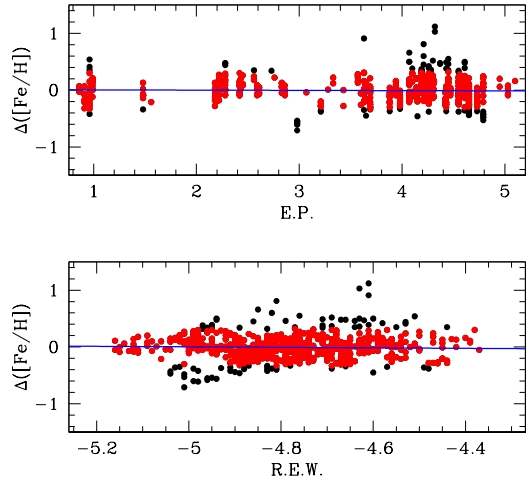}
\caption{$\Delta$[Fe/H] abundances vs. Excitation Potential (E.P., upper panel) and vs. 
Reduced Equivalent Width (R.E.W., lower panel) for all RGB and RC stars.
Rejected points are indicated by black circles. See text for more details}
\label{fig2}
\end{figure}

The abundance analysis was performed using ATLAS9 atmospheric models
\citep{Ku70} and the Local Thermodynamic Equilibrium (LTE) program MOOG \citep{Sn73}.
SiI, CaI, TiI, TiII, CrI, FeI, FeII, and NiI abundances were estimated
using the equivalent width (EQW) method. EQWs were measured manually
adjusting a gaussian to each spectral line. Lines affected by blending or
telluric contamination were rejected. The main problem in this case was the continuum
determination, due to the very high metallicity. We solved this by comparing
our spectra with synthetics calculated using the same atmospheric parameters as the
targets and adopting as continuum only those portions of the observed spectra
where the corresponding synthetic was $\leq$1\% below the theoretical continuum. 
NaI, MgI, AlI, YII,  LaII, and EuII abundances were obtained using the
spectrum synthesis method. For this purpose, 5 synthetic spectra were generated
for each line with 0.25 dex abundance steps inbetween them and then compared with the observed
spectrum. The  line-list and the methodology we used are the same used in
previous papers (e.g. \citealt{Vi13,Gei12}), so we refer to those articles for a
detailed discussion of this particular point. 
Here we emphasize the fact that we took hyperfine splitting into account for Ba
as in our previous studies. This is particularly important because Ba
lines are very strong and hyperfine splitting helps to remove the line-core
saturation, producing a change in the final abundance as  estimated by the
spectrum synthesis method by up to 0.1 dex. Also Y and Eu 
are affected by hyperfine splitting, but their lines are much weaker compared
to Ba and the line-core saturation is negligible.
On the other hand, Na is an element affected by NLTE effects. For this reason we checked for NLTE corrections using the INSPEC \footnote{version 1.0 (http://inspect.coolstars19.com)} database. The corrections turned out to be very similar for all the stars, in the range -0.10$\div$-0.14 dex. We decided to apply a common correction of -0.12 dex to all the targets.

At odds with \citet{Gei12}, here we could not measure [O/Fe] because the oxygen
line at 6300 \AA\ was too badly blended with the oxygen atmospheric emission line that
falls exactly at the center of the stellar line given the geocentric radial velocity at
the time of our observations.

As initial atmospheric parameters, we used the values reported in
\citet{Gei12}. Here we just want to remember that those parameters are based on multicolor 
photometry. T$_{\rm eff}$ was derived from B-V, V-I, V-J, V-H, V-K, J-H, and J-K colors
adopting a reddening of E(B-V)=0.13.
Surface gravities (log(g)) were obtained from the canonical equation:
$$ \log\left(\frac{g}{g_{\odot}}\right) =
         \log\left(\frac{M}{M_{\odot}}\right)
         + 4 \log\left(\frac{T_{\rm{eff}}}{T_{\odot}}\right)
         - \log\left(\frac{L}{L_{\odot}}\right). $$
where the mass M was assumed to be 1.13 and 1.05 M$_{\odot}$ for RGB and
RC/AGB stars respectively. Luminosity L/L$_{\odot}$ was obtained from the absolute magnitude M$_{\rm V}$ assuming an apparent distance modulus of (m-M)$_{\rm V}$=13.44.
Finally, the micro-turbulent velocity (v$_{\rm t}$) was obtained from the
relation of \citet{Gra96} that takes both temperature and gravity into account.

Since we used high-quality photometry and seven color combinations for the temperature determination, the random errors
are very low.  The temperature error was obtained by comparing the individual
color-based determinations for each star, while the errors in gravity and
micro-turbulence were obtained applying error propagation to the previous
equations assuming an internal mass uncertainty of 0.05 M$_{\odot}$. We obtained
$\sigma_{T_{eff}}$=10 K, $\sigma_{log(g)}$=0.05 dex, and $\sigma_{v_{t}}$=0.04 km/s.
With respect to our previous analysis, here we have spectra with much higher
resolution and many more Fe lines (65-75 per star), so the error due to 
the S/N is very low, of the order of $\sigma_{[Fe/H]}$0.01 dex.

Having so many available Fe lines, we also tried to derive the atmospheric parameters spectroscopically, where
T$_{\rm eff}$, log(g), and v$_{\rm t}$ were re-adjusted and new 
atmospheric models calculated in an interactive way in order to remove trends 
in excitation potential and reduced equivalent width  versus 
abundance for T$_{\rm eff}$ and v$_{\rm t}$, respectively, 
and to satisfy the ionization equilibrium between FeI and FeII for log(g). In
this case the [Fe/H] value of the model was changed at each iteration according
to the output of the previous abundance analysis. However, the final result was not as
accurate as that based on photometry because errors in temperature and
micro-turbulence were much higher ($\sim$50$-$60 K and $\sim$0.10$-$0.15
km/s respectively). That is because in this line by line analysis the
spread of results from the many Fe lines was relatively large and so it was very difficult to remove outliers for a proper spectroscopic parameter determination.
We conclude that the use of photometric-based parameters was the best way
to pursue our analysis in order to have the smallest internal error.

However, photometric-based parameters are affected by systematic errors since
we assumed a reddening and a distance modulus that are uncertain. The micro-turbulence scale we used was also based on an equation
that could contain systematics. In order to remove such systematics as much as
possible, we used the same spectroscopic analysis described above, but with
a variation. First of all we put all the single FeI/II and TiI/II abundances of all the stars together. For this purpose we calculated normalized abundances ($\Delta$[El./H)],
where we subtracted the mean abundance  of the star to the FeI/II and TiI/II abundances obtained from each single line.
Then we applied to the photometric-based T$_{\rm eff}$, log(g), and v$_{\rm t}$
scales three zero-point corrections ($\Delta$T$_{eff}$,$\Delta$log(g), and
$\Delta$v$_{t}$) in order to remove trends in excitation potential and reduced
equivalent width versus abundance (for the temperature and the micro-turbulence
scales respectively), and to satisfy the ionization equilibrium between FeI and FeII,
and TiI and TiII simultaneously (for the log(g) scale). 
Since our targets cover different evolutionary phases, we decided to
divide them in three groups: RC stars, upper RGB stars, and lower RGB
stars. We applied the procedure described above to the three groups
separately. RC and upper RGB stars turned out to require the same zero
point-corrections, while for lower RGB stars the zero-point correction for
microturbulence is slightly lower.
The result is reported in Fig.\ref{fig2} for RC and upper RGB stars together.
The zero-point corrections we had to apply are the following:
$\Delta$T$_{eff}$=-25 K and $\Delta$log(g)=-0.30 dex for all the groups, while
$\Delta$v$_{t}$=+0.13 km/s for RC and upper RGB stars, and
$\Delta$v$_{t}$=+0.08 km/s for lower RGB stars. 
We also applied a sigma-clipping rejection method. Rejected abundances are in
black, while good abundances are in red. 

A great advantage of this method is that it allows identification of 
outliers (i.e. those lines that for blending or other reasons give relatively extreme
abundances) efficiently, which can then be easily removed. In this way the final abundances are greatly improved. We applied the same outlier removing process also to the other elements measured by EQWs. We do not show these plots here but they are similar to
Fig.\ref{fig2}. The final abundances of Si, Ca, Ti, Cr, Fe, and Ni were
calculated using only the lines with good abundances left after the sigma-clipping
rejection. For Ti, we give the mean of the TiI and TiII abundances. The
results of the abundance analysis are reported in Tab.~\ref{t2}.

As a final comment, we underline the fact that micro-turbulence is a critical
parameter because of the high-metallicity and the relatively low-temperature
of our stars, and final abundances depend strongly on it. This is the first
time that this parameter is obtained directly from the spectra of NGC 6791 stars
and not assumed from some equation. This makes us confident that the final
[Fe/H] values we give are as robust as possible.

Error analysis has been conducted assuming star $\#18243$ as
representative of the sample. We varied its T$_{\rm eff}$, log(g), [Fe/H], and v$_{\rm t}$ according to the internal atmospheric errors reported above, and redetermined the abundances.
We also performed an error analysis assuming more conservative errors on the parameters, that is $\sigma_{T_{eff}}$=50 K, $\sigma_{log(g)}$=0.20 dex, $\sigma_{v_{t}}$=0.10 km/s, and $\sigma_{[Fe/H]}$=0.05 dex
Results are shown in Tab.~\ref{t3}, including the error due to the noise
of the spectra. Errors obtained using the more conservative errors on the parameters are those within brackets.
Error due to the noise was obtained for elements whose abundance was
obtained by EQWs, as the errors on the mean given by
MOOG, and, for elements whose abundance was obtained by spectrum-synthesis, as
the error given by the fitting procedure. $\Delta_{\rm tot}$ is the square root of the
sum of the squares of the individual errors. In Tab.~\ref{t3} for each element we report the
observed spread of the sample ($RMS_{\rm obs}$) with its error and in the final column the
significance (in units of $ \sigma$) calculated as the absolute value of the difference between 
$RMS_{\rm obs}$ and $\Delta_{\rm tot}$ divided by the error on $RMS_{\rm  obs}$. 
This tells us if the observed dispersion $RMS_{\rm obs}$ is
intrinsic or due to observational errors. Values larger than 3$\sigma$ imply
an intrinsic dispersion in the species chemical abundance among
the cluster stars. Again, values within brackets are those calculated using the more conservative errors on the parameters.

\begin{figure}
\plotone{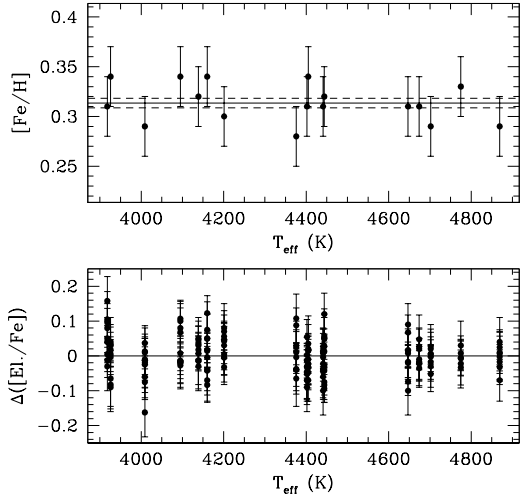}
\caption{Upper panel: [Fe/H] abundances vs. temperature for our sample. The mean value (continuous line) and the 1$\sigma$ 
error on the mean (dashed lines) are indicated.
Lower panel: Normalized $\Delta$[El./Fe] vs. temperature. The zero slope trend is indicated as a continuous line. No trend is
visible. See text for more details.}
\label{fig3}
\end{figure}

We performed a further check on the internal consistency of our results
by plotting in the upper panel of Fig.~\ref{fig3} the [Fe/H] abundances of
our stars as a function of temperature. No trend is present. For the other
elements a similar plot can be misleading since their abundances are not as
accurate as those for iron and can deviate significantly from the mean value
of the cluster creating a false trend. This is because their abundances  are
based on fewer spectral lines than Fe, implying that some outliers can still
be present in spite of the procedure we applied to remove them as much as
possible. Because of this, a global plot is more significant. For this purpose,
first of all we considered each element separately and subtracted from the
abundance of each star the mean value of the cluster, obtaining what we call normalized
abundance ratios ($\Delta$[El./Fe]). Then we plot all the normalized abundance ratios together as a function of temperature in the lower panel of Fig.~\ref{fig3}. The
advantage of this procedure is that we have a much larger sample and
abundance ratios that deviate significantly from the mean value of the
cluster have a much lower impact on the final trend. Fig.~\ref{fig3} reveals
that in fact  no temperature trend is present for our normalized abundances either.

\begin{figure}
\plotone{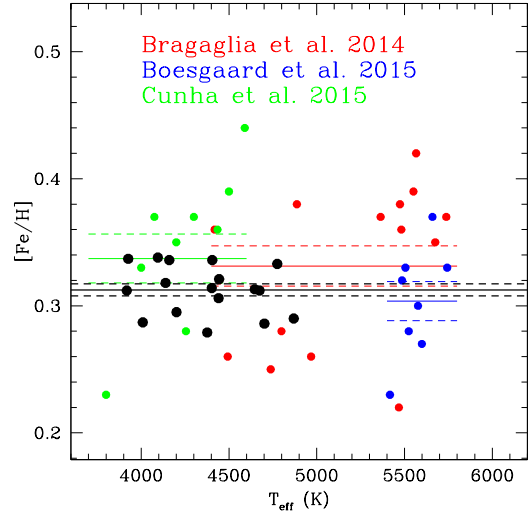}
\caption{[Fe/H] vs. temperature relation for our data (black points) and 
three recent spectroscopic studies. Continuous
lines are the mean values, while dashed lines are the 1$\sigma$ error on the mean.}
\label{fig4}
\end{figure}

\begin{figure}
\plotone{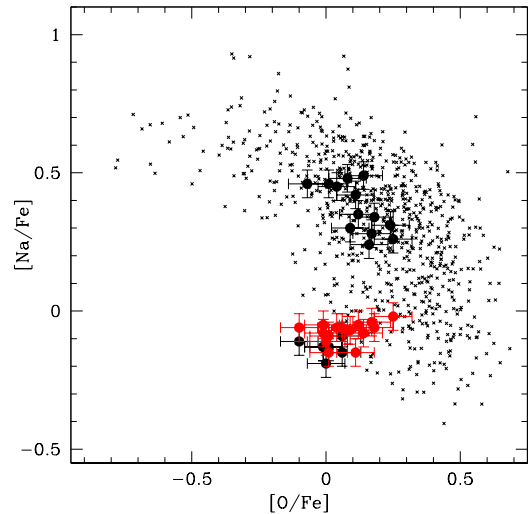}
\caption{Na-O anticorrelation trend for globular clusters from
  \citet[crosses]{Car09} as a reference. For NGC 6791 we indicate in black the measurements from \citet{Gei12}, and in red our new determinations.}
\label{fig05}
\end{figure}

\section{Results}

The mean iron content we obtained is:

$$[Fe/H]=+0.313\pm0.003$$

with a dispersion of:

$$\sigma_{[Fe/H]}=0.020\pm0.003$$

Reported errors are errors on the mean. This value is lower than our previous measurement
of $[Fe/H]=+0.42\pm0.01$ \citep{Gei12}, but well in line with recent determinations.
The difference with respect \citet{Gei12} is mainly due to the different microturbulence
scale we adopted here.
\citet{Bra14} finds $[Fe/H]=+0.34\pm0.02$, \citet{Boe15} derives $+0.30\pm0.02$, \citet{Cu15} $+0.34\pm0.06$, and \citet{Lin17} $+0.31\pm0.01$.
The latter three studies all use the same APOGEE dataset. Our results agree
nicely with \citet{Boe15} and Linden et al. within 1$\sigma$, while both \citet{Bra14} and \citet{Cu15} have a slightly higher metallicity. However, all values are in agreement at the 2$\sigma$ level. In Fig.~\ref{fig4} we report our
data (black points) with the results from \citet[red points]{Bra14},
\citet[blue points]{Boe15}, and \citet[green points]{Cu15}.  We notice
that both \citet{Bra14} and \citet{Cu15} Fe abundances have a possible trend with
temperature. 

While our metallicity is not as extreme as some past measurements (most notably the super-metal-rich value of +0.75 reported by \citealt{Spi71}), this metallicity reconfirms NGC 6791 as possibly the most metal rich open cluster known in the Galaxy, and the only one with such an extreme combination of age and metallicity.
The measured iron dispersion in Tab.~\ref{t3} agrees well with the expected
dispersion due to measurement errors so we have no evidence for an intrinsic Fe abundance
spread, as expected. 

Al shows a super-solar value of +0.24 dex, that is larger than any Thin or Thick Disk star
at the same metallicity. The same behavior is shared by Mg (0.15 dex) and Ti (0.13), although not so extreme. The other two $\alpha$ elements Si and Ca are on average solar-scaled (Si) or slightly sub-solar (Ca). The mean $alpha$-element content of NGC 6791
based on Mg, Si, Ca, and Ti is solar-scaled within the errors:

$$[\alpha/Fe]=+0.06\pm0.05$$

As far as iron-peak and heavy elements are concerned, Cr, Ni, Y, and Eu are super-solar, 
while Ba is solar scaled.

\subsection{About the Na spread}

\begin{figure}
\plotone{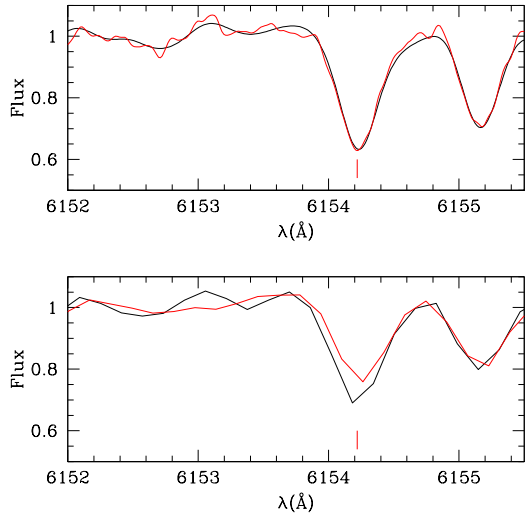}
\caption{Comparison of UVES spectra (upper panel) and Hydra spectra (lower panel) for the very similar RC stars \#T05 (black lines) and \#T18 (red lines). The Na line is indicated. In UVES data we find no evidence for any absorption strength variation at odd with what we we see in Hydra data. See text for more details} 
\label{fig6}
\end{figure}

One of the chief aims of this paper is to confirm or disprove the very surprising result found in \citet{Gei12}.  On the basis of their Na abundances, they suggested the
presence of an intrinsic Na spread and even a slight Na-O anti-correlation, which led to the conclusion that NGC 6791 was the least massive star cluster hosting multiple stellar populations and the first open cluster to display this behavior.
In this study, we purposely re-observed the same brightest \citet{Gei12} stars, previously observed with Hydra at Kitt Peak, but now using UVES at much higher resolution and much wider wavelength coverage. We compare the present results (red points) with \citet[red points]{Gei12} in Fig.~\ref{fig05}. We do not have Oxigen here, however we assume the same
\citet{Gei12} values since we calculated that the change in the atmospheric parameters
affect only marginally (-0.01$\div$-0.02 dex) the [O/Fe] values we published there. 
Fig.~\ref{fig05} reveals that our current data do not support a Na spread anymore. We investigated possible reasons for this discrepancy  and identify the
source as most likely due to a reduction problem of the Hydra spectra in
\citet{Gei12}. Fig.~\ref{fig6} illustrates this evidence. Red lines are the two
spectra of the star \#T18, while black lines are the two spectra of the star
\#T05. These two RC stars have the same atmospheric parameters, so any difference
in the strength of a given spectral line directly implies a difference in the abundance
of the element that produces the line. Upper panel shows the current UVES
data, while the lower panel shows the old Hydra data. The Na line at 6154 \AA\ is
indicated. In the Hydra data, the Na lines of the two stars have different
strengths, leading to different [Na/Fe] values. On the other hand, UVES data
show that the Na lines have the same strength, implying  the same Na abundance
for the two stars. This means that the Na abundance determinations obtained
from the Hydra spectra were likely affected by some kind of instrumental problem, very
likely a bad flat-field correction or a bad pixel. We remember here that in Hydra data
we had only the Na line at 6140\AA available (the line at 6160 \AA was too heavily blended) and that the line sampling was not optimal, making the identification of bad pixel problematic.
We conclude that NGC~6791 does not host a Na abundance spread and therefore does not display any evidence for multiple stellar populations. This conclusion is supported also by the other light elements Mg and Al that, according to table \ref{t3}, do not show an intrinsic spread. Indeed, there is no evidence for a real spread in any of the 12 elements we measure. This is in accord with the findings of \citet{Bra14,Boe15,Cu15}, but our data have the smallest errors. 

Have we thus definitively solved at least one of the mysteries surrounding NGC 6791? But
what about the CN spreads seen in previous low resolution studies by \citet{Huf95} and \citet{Car12}? Note that \citet{Bob16} reanalyzed the SEGUE spectra studied by Carerra and found no strong evidence for any CN (or CH) variations. We cannot derive C, N or O abundances from our data. But the available APOGEE data and analyses do not show any evidence for intrinsic variation in any of these elements \citep{Cu15}. Additional APOGEE data has been obtained to further address this issue but it appears that NGC 6791, despite our previous claim, does not in fact host multiple populations.

\section{Discussion and Conclusions}

Recent studies, both observational and theoretical,  have addressed once again
the issue of the origin of NGC 6791  and its association with the various Galactic
components: Thin Disk, Thick Disk, or Bulge.
On the observational side, the comparison of NGC 6791 elemental abundances and
abundance ratios with the DR13 release of APOGEE data led  \citet{Lin17}
to suggest that NGC 6791 is a member of the Galactic Thick Disk. Their
arguments proceed along three levels. First of all, the comparison of the
metallicity and $\alpha$-ratio seems to suggest a similarity between NGC 6791
and the high metallicity, high $\alpha$-ratio tail of the Galactic Thick
Disk. Second, the earlier suggestion \citep{Jil12} that NGC 6791 might have
formed close to the Bulge is ruled out  by the difficulty to displace
such a massive cluster to its actual position. And third, the actual cluster
location at 1 kpc above the Galactic plane makes it difficult to envisage a
possible connection with the Galactic Thin Disk. 
On the theoretical side, \citet{Mar17} provide an independent
argument that NGC 6791 might indeed have formed close to the Bulge, in the inner 3-5kpc of the Galaxy, and then suffered radial migration and was 
displaced to where we observe it today.
Based on our new, high quality data  presented in this work, we now reconsider the various arguments in an attempt to provide a more observationally robust scenario for NGC 6791's origin. 

\begin{figure}
\plotone{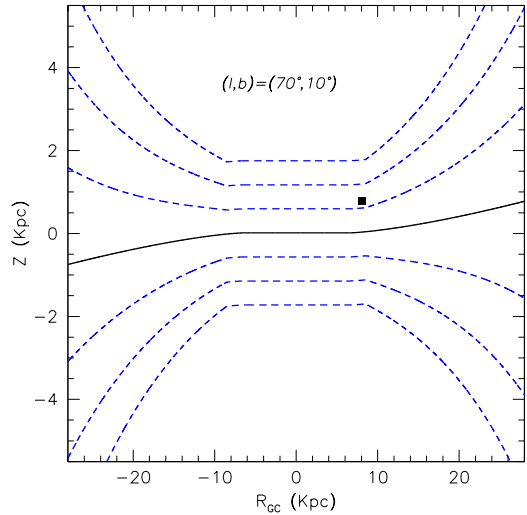}
\caption{Location in the X,Z plane of NGC 6791 with respect the warped and flared Galactic disk. See text for more details}
\label{fig07}
\end{figure}

\begin{figure*}
\plotone{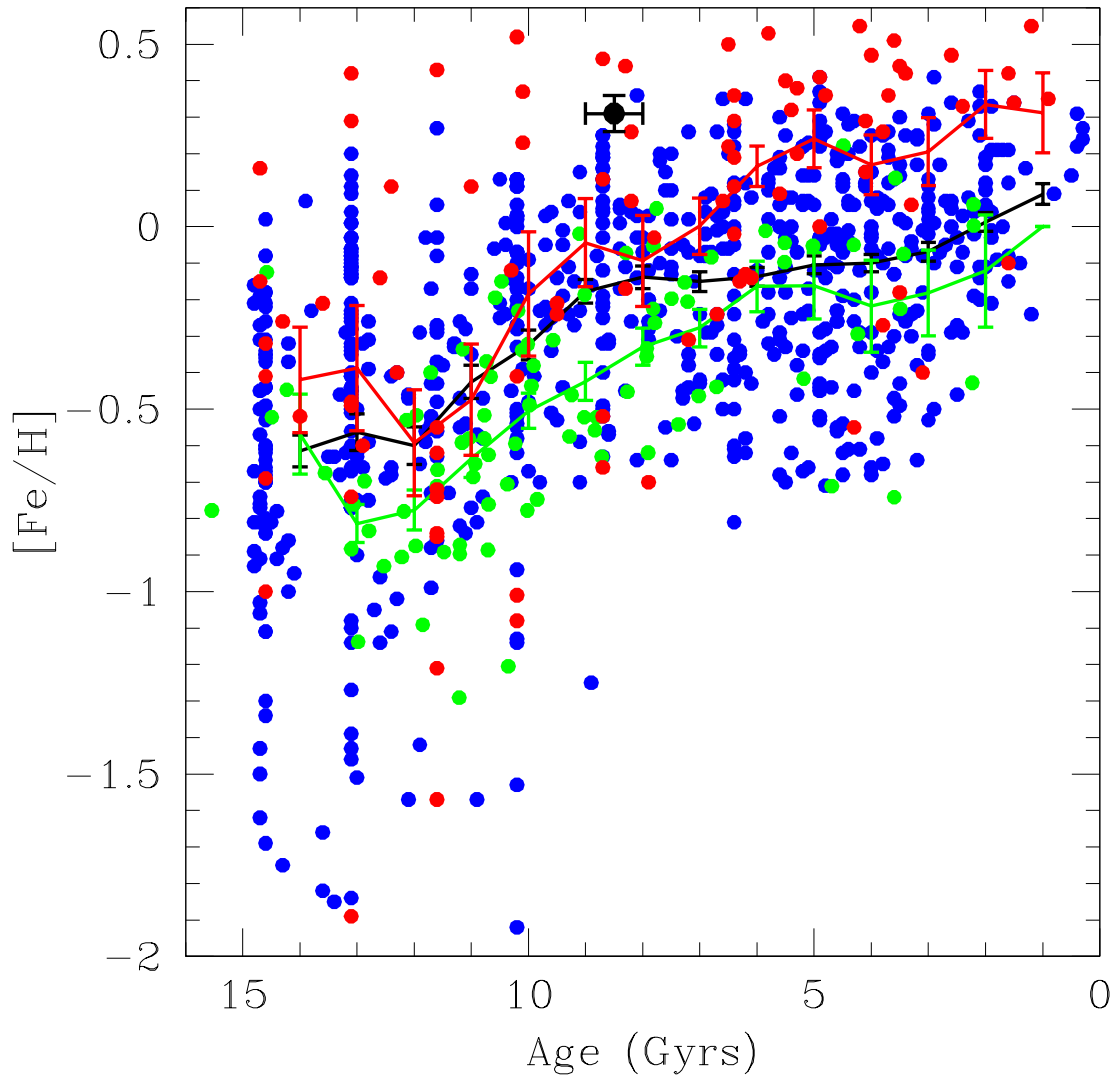}
\caption{Age-metallicity relations for the Thin Disk (blue points), Thick Disk
(green points), and Bulge (red points). Continuous lines with error bars are
the respective mean relations. The mean relation for the Thin Disk is shown
in black to be more visible. We assumed an error bar of 0.05 for the NGC~6791
iron content to be conservative (black point).}
\label{fig08}
\end{figure*}

First of all, we discuss the possibility that NGC~6791 belongs to
the Thin Disk instead of the Thick Disk based on its position only. If NGC 6791 was a Thick Disk object, it would be the only open cluster associated with
this Galactic structure, with the possible addition of Gaia 1, which has been recently associated to the Thick Disk \citep{Ko18} on the basis of the very same \citet{Lin17}
argument, namely that its location is too high to be compatible with the
Galactic Thin Disk. This argument, however, is embarrassingly weak. A wealth
of observational data have been accumulated over the last 10 years that
indicate how the Galactic Disk, both Thin and Thick, is not a plain flat
structure, but possesses a significant warp and flare both in its gaseous
and stellar components, and both in its young and old populations. These data
are however disappointingly neglected.
The case of Gaia 1 is easy to accomodate \citep{Car07,Car18} since it is an outer Disk object and the outer Thin Disk has been repeatedly shown to be significantly
warped and to harbour a number of intermediate age and old open clusters, to
which Gaia 1 bears much resemblance. There are other open clusters presently located more than 1 kpc above or below the formal Galactic plane ($b=0^o$) \citep{Chen03}. However, as in the case of Gaia 1, they are all located in the outer part of the Galaxy and therefore they very likely belong to the warped and flared Thin Disk \citep{Mom05,Car07}.
The case of NGC 6791 seems more difficult to sort out. In reality, this
difficulty is simply apparent, because a quick inspection of the warped
structure of the Disk convincingly shows that at the distance and location of
NGC 6791 the disk is actually about 1 kpc off the formal b=0$^0$ Galactic plane. This
is shown in Fig.~\ref{fig07}, where the Galactic Disk as traced by red clump stars is shown
for scale heights of 0 (solid line), 1,2 and 3 (dashed lines) in the direction of NGC 6791.
From this one can easily infer that NGC 6791 comfortably sits at just over 1 RC scale
height from the formal Galactic plane. Since clump stars are genuine
population I objects and trace the Galactic Thin Disk, the conclusion can be
easily drawn that NGC 6791 can also be spatially a member of the Galactic Thin Disk.

\begin{figure*}
\plotone{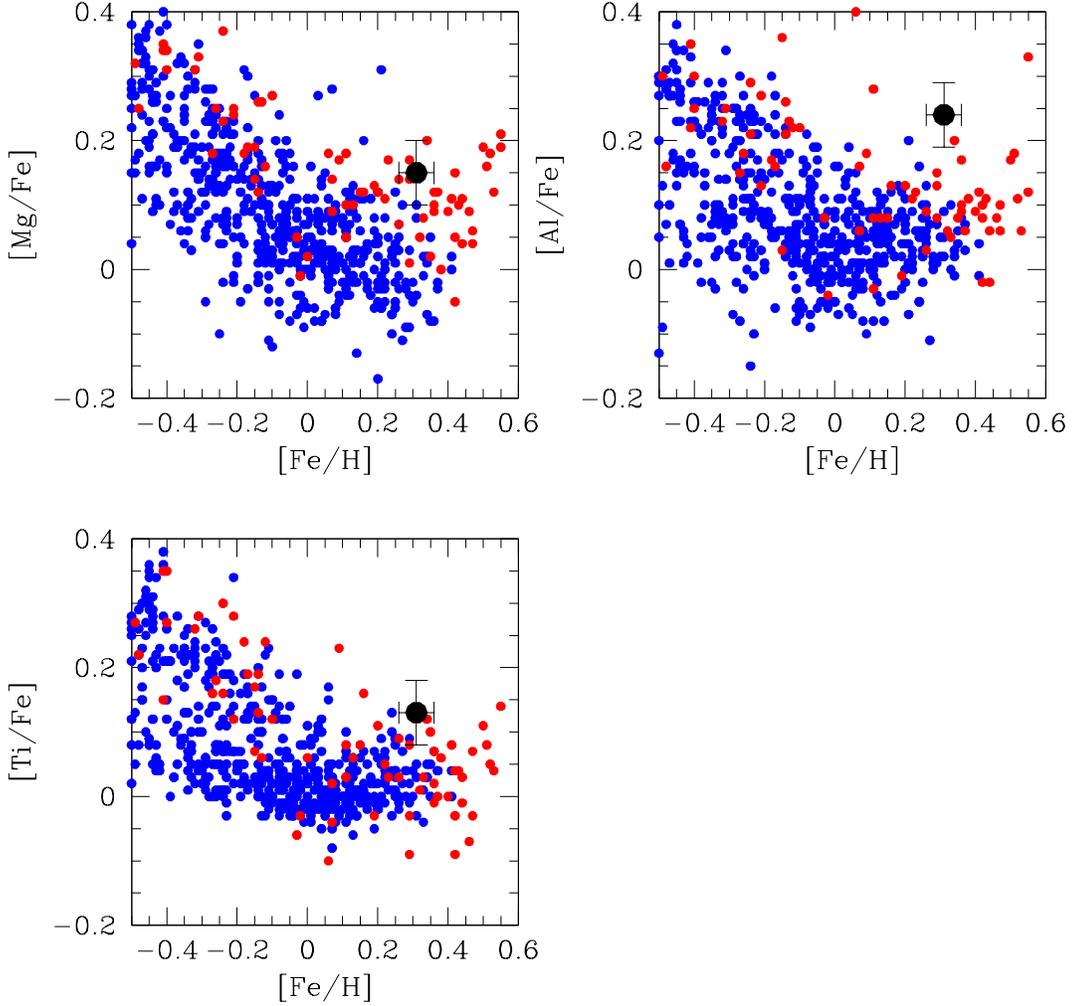}
\caption{[Mg/Fe], [Al/Fe], and [Ti/Fe] abundances as a function of [Fe/H] for the Thin
Disk (blue points) and the Bulge (red points). NGC~6791 is the black point
with error bars. We assumed a conservative error of 0.05 for all elements.}
\label{fig09}
\end{figure*}

\begin{figure*}
\plotone{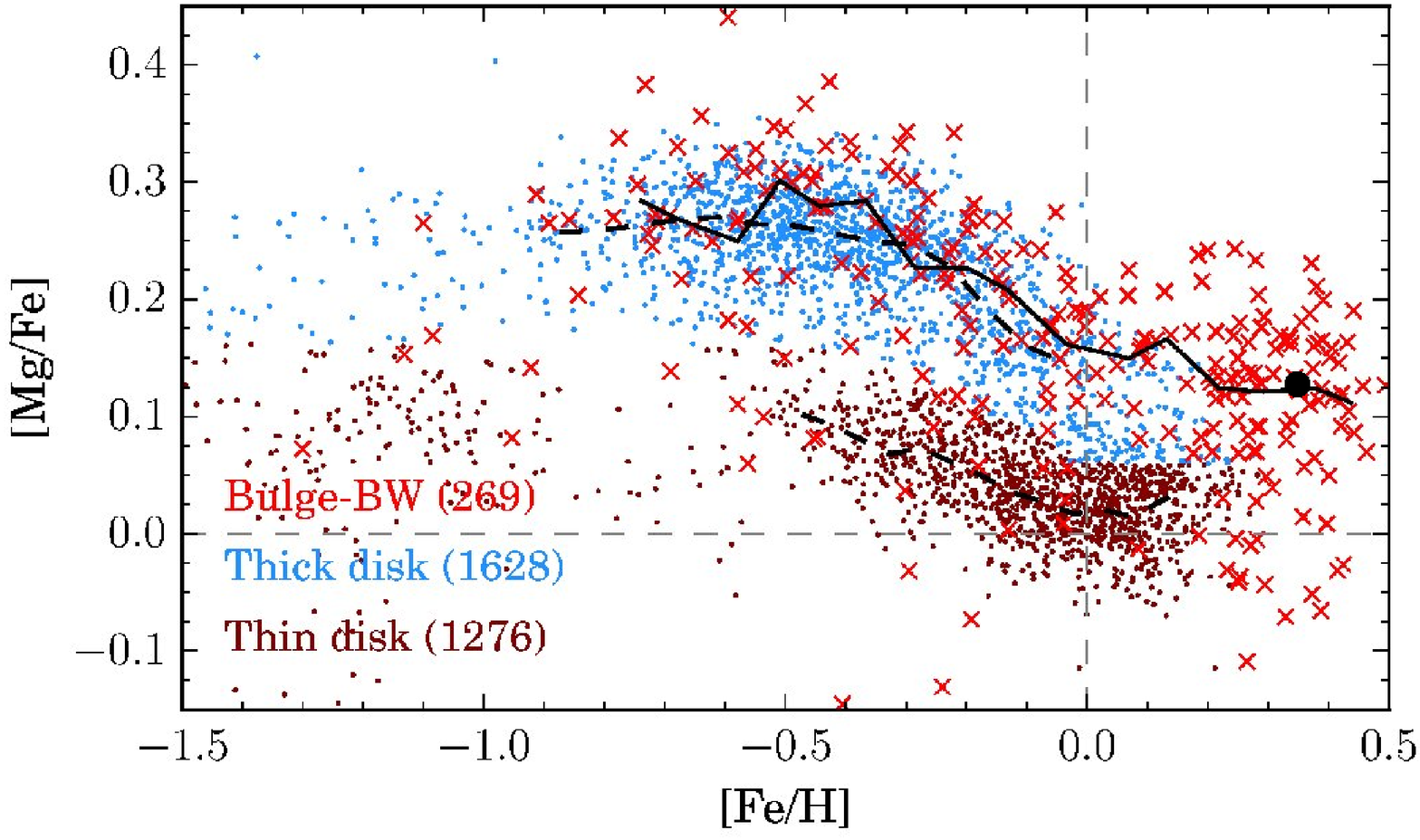}
\caption{[Mg/Fe] vs. [Fe/H] abundances for Thin Disk (blue points), Thick Disk (brown points), and Bulge (red crosses) star from \citet{Lin17}. NGC 6791 is represented by the black circle.}
\label{fig10}
\end{figure*}

Moving to chemical arguments, we note that the cluster metallicity estimates
have been decreasing over the years until recently, and nowadays there is a
general consensus that NGC 6791 has [Fe/H] around +0.3 dex. This lower value
comes as a consequence of better quality data and better analysis of the
stellar atmospheres in the high metallicity regime. At [Fe/H] around +0.3, NGC
6791 can be associated either with one of the peaks in the Bulge metallicity distribution \citep{Gar18}, or with the extreme tails of either the Thick or
Thin Disks.  On purely statistical grounds, it is certainly more likely that it is a typical member of its Galactic component than an extreme one, and therefore a Bulge origin is favored. We found [Fe/H]=+0.313$\pm$0.005, a value which agrees
with the most recent estimates.
We can combine this metallicity with the age of the cluster,
\citet[8.5$\pm$0.5 Gyrs]{Buz12}, and compare these data with the
age-metallicity relations of the Thin and the Thick Disks and the Bulge. 
For this purpose, we used the results of \citet{Ben07,Ben14,Ben17}. 
In Fig.\ref{fig08} blue points represent the relation for the
Thin Disk, green points the relation for the Thick Disk, and red points the
relation for the Bulge. Black, green, and red continuous lines with errorbars
are the mean relations as obtained from the respective data. We used a black
line for the Thin Disk data (blue points) to make it more visible.
NGC~6791 (the black point with errorbars) is clearly above all the mean relations.
However, note that it lies  well above ANY member of the Thick Disk. This effectively rules it out as a member of this Galactic component. It is also only barely
compatible with the Thin Disk since it lies on the very upper edge of the area
covered by Thin Disk stars. On the other hand, it lies comfortably within the area
covered by Bulge stars. Indeed, a few of them are as old or older than NGC~6791 and
even more metal rich. We conclude that, according to this analysis, NGC~6791
most likely belongs to the Bulge, although we cannot completely rule out its
membership in the Thin Disk.

Chemistry provides us with further strong evidence for its likely bulge
nature. In Fig.~\ref{fig09} we plot [Mg/Fe], [Al/Fe], and [Ti/Fe] abundances
as a function of [Fe/H] for the Thin Disk (blue points) and the Bulge (red
points) from \citet{Ben14,Ben17}.
We again see that, as far as Mg and Ti are concerned, NGC~6791 is fully
compatible with the Bulge, although a possible relation with the Thin Disk
cannot be ruled out, however with a very low probability. In the case of Al,
it is very hard to reconcile NGC~6791 with the Thin Disk, and the association
with the Bulge is left as the only possible hypothesis. The only week
point of this comparison is that our results are not homogeneous with those
from \citet{Ben14,Ben17}.
For this reason we look at a totally independent and even larger but
still homogeneous dataset, that of \citet{Sch17} using 1276 Thin disk stars,
1628 Thick disk stars and 269 Bulge stars from the APOGEE survey. Adding NGC
6791 [Fe/H] and [Mg/Fe] values from APOGEE data ([Fe/H]=+0.34 and
[Mg/Fe]=+0.13, see \citet{Lin17}) to their plot of [Mg/Fe] vs. [Fe/H], we
find the cluster resides in a region where ONLY bulge stars are found. The
nearest Thin or Thick Disk stars lie many $\sigma$ away. We report this
comparing in Fig.~\ref{fig10}, where NGC 6791 is represented by the black
circle.

This result is further corroborated by yet a second, albeit much smaller
dataset - that of \citet{Jon16}, where they analyze FLAMES data for a sample
of local Thick Disk stars vs. Bulge stars. Our NGC 6791 data place it along
the trend for bulge stars in the same [Mg/Fe] vs. [Fe/H] plane very far away
from any Thick Disk stars.

The preponderance of the chemical evidence is unequivocal: NGC~6791 is very
likely a cluster that was born in the Galactic Bulge. The age-metallicity
diagram also supports this interpretation. Any possible association with the
Galactic Thick or thin Disks is essentially ruled out. Nevertheless, as we
argue below, although probabilities are small, given a large enough sample
outliers do occur and NGC 6791 if nothing else has proven to be an exceptional
exception to the rules.

We are left with the conundrum of explaining how an object originating in the Bulge has
managed to move outwards by at least 5kpc, about a factor of 2 in Galactocentric distance, during its lifetime. 
The problem is exacerbated by the fact that this is not a single star but a
massive object, making it less susceptible to effects that would otherwise be
quite effective on single stars. Several studies have investigated this
scenario dynamically, most recently  by \citet{Mar17}, who refined and
strengthened earlier suggestions by \citet{Jil12} and \citet{Da15}. They
investigated in detail the possibility that NGC 6791 formed in the Inner Disk
or Bulge and has radially migrated to its current position. Given its high
metallicity and what we know of the Thin Disk and Bulge metallicity
distributions as a function of Galactocentric distance (e.g. \citealt{Gar18}),
it is likely that it formed at a Galactocentric distance of between 3 - 5kpc,
and has therefore moved outwards by 3 - 5kpc over its lifetime. \citet{Mar17}
find only a 0.1$\%$ probability that this actually happened, given all we know
about the cluster and the Galactic potential and dynamics. This is in
reasonable agreement with the \citet{Jil12} probability of 0.4$\%$. However,
as \citet{Mar17} point out, this means we only need to have started with a few
hundred to a thousand such clusters to find one today that actually achieved
this feat. They also find that, in order to survive such radial migration over
its lifetime, the original NGC 6791 must have been much more massive, about an
order of magnitude. \citet{Da15} have indeed uncovered evidence for tidal
tails and mass loss from NGC 6791, and estimate its original mass could have
been $\geq 10^5 M_\odot$, more than an order of magnitude larger than its
current mass of $5\times 10^3 M_\odot$. \citet{Mar17} finally conclude
that such a cluster born at a Galactocentric distance between 3 - 5kpc 8 Gyr
ago would have a 0.2$\%$ probability of being found today where it actually
is.

Our observational results combine with these simulations to paint a convincing
scenario in which NGC 6791 almost certainly must have formed in the Bulge or
inner Disk chemically and has had a slight but non-negligible chance to
radially migrate to its current location and orbit dynamically. Thus, this
overall scenario of NGC 6791's formation in the inner Galaxy appears  very
appealing.
\ \\

\acknowledgments
SV and PA gratefully acknowledge the support provided by Fondecyt
reg. n. 1170518. D.G. gratefully acknowledges support from the Chilean BASAL
Centro de Excelencia en Astrofisica y Tecnologias Afines (CATA) grant
PFB-06/2007. D.G. also acknowledges financial support from the Dirección de
Investigación y Desarrollo de la Universidad de La Serena through the Programa
de Incentivo a la Investigación de Académicos (PIA-DIDULS). 

\vspace{5mm}
\facilities{VLT:FLAMES}

\software{MOOG \citep{Sn73}}

\end{document}